\documentstyle[preprint,aps]{revtex}
\begin{document}
\draft
\title{\rightline{\rm \normalsize TIPAC-930010}\vspace*{2.0cm}
Fermion and Anti-Fermion Effective
Masses in High Temperature Gauge Theories in $CP$-Asymmetric Background}
\author{A. Erdas}
\address{
Dipartimento di Scienze Fisiche, Universit\`a di Cagliari,
09124 Cagliari, Italy\\
I.N.F.N. Sezione di Cagliari, 09127 Cagliari, Italy}
\author{C. W. Kim and J. A. Lee}
\address{
Department of Physics and Astronomy \\
The Johns Hopkins University, Baltimore, MD 21218
\vspace*{1.5cm}}

\maketitle
\begin {abstract} We calculate the splitting between
fermion and anti-fermion effective masses in high temperature gauge
theories in the presence of a non-vanishing chemical potential due to
the $CP$-asymmetric fermionic background. In particular we consider the case
of left-handed leptons in the
$SU(2)\otimes U(1)$ theory
when the temperature is above $250$ GeV and the gauge symmetry is
restored.
\vspace*{-0.3truecm}

\noindent \underline{$\phantom{space space space space}$}

\noindent
 erdas@vaxca2.unica.it; cwkim@jhuvms.hcf.jhu.edu; leej@dirac.pha.jhu.edu
\end {abstract}
\newpage
\section{Introduction}
\label{intro}

The behavior of gauge theories at finite temperature and density
has been extensively investigated in the literature. In particular, the
modifications to the fermion dispersion relation due to high-temperature
effects for a chirally invariant gauge theory have been studied by
Weldon \cite{Weldon} who demonstrated that right-handed and left-handed
leptons in the $SU(2)\otimes U(1)$ model acquire different effective masses.
The properties of neutrinos at low temperature and density have
also been studied by many authors \cite{Notzold-Raffelt,Masood,Nieves}.

In this article we examine the properties of chirally invariant gauge
theories at high temperature, in particular in the presence of a
non-vanishing chemical
potential due to a CP-asymmetric fermionic background;
one of these theories can be, for example, the Standard Electroweak
Model at temperatures for which the $SU(2)\otimes U(1)$ symmetry is
restored.
The symmetry restoration that occurs in the Standard Electroweak Model
at high enough temperature has been studied
extensively \cite{Sher}:
above $T=250$GeV the Higgs vacuum expectation value vanishes and
so do the masses of the gauge bosons, scalars and the fermions.

The interaction of a particle with the heat bath changes the dispersion
relation from the tree level relation.
Weldon \cite{Weldon} calculated this temperature correction
for chirally symmetric gauge theories in
a CP-symmetric medium at high temperature.
The purpose of this paper is to extend the Weldon's work to include
the case of a non-vanishing chemical potential for the fermions.
In the scenario in which the baryogenesis takes place right after the GUTS
phase transition and $B-L$ conservation is assumed,
the number of particles in the $SU(2)\otimes U(1)$ symmetric
phase of the early universe could be slightly
larger than the number
of antiparticles, implying non-vanishing chemical potentials
and leading to the current particle-antiparticle
asymmetry \cite{Dine,Cohen}.

To compute the effective fermion masses under the above condition,
we need to evaluate the fermion self-energy \cite{Weldon}
$$\Sigma(K)=-a\gamma_\mu K^{\mu} -b\gamma_\mu u^{\mu}\, ,
\eqno(1.1)$$
where $u^\alpha$ is the four-velocity of the heat bath,
$K^\alpha$ is the four-momentum of a fermion under consideration
and $a$ and $b$
are Lorentz-invariant functions that depend on the Lorentz-scalars
 defined by\cite{Weldon}
$$\omega\equiv K^\alpha u_\alpha,\;\;\;\;\; k\equiv \left[(K^\alpha u_\alpha
)^2-K^2\right]^{1/2}\,.
\eqno(1.2)$$
We then find the poles of the fermion propagator
$$S(K)=[(1+a)\gamma_\mu K^{\mu}+b\gamma_\mu u^{\mu}]^{-1}\,.
\eqno(1.3)$$

In Section II we compute the fermion self-energy in the presence of
a non-vanishing chemical potential.
Using the results of Section II, we  find, in Section III, an
integral equation for the effective mass in the presence of
the chemical potential
$\mu\ll T$ and use this integral equation to calculate
the mass-splitting between fermions and anti-fermions. Finally in Section
IV we discuss some implications of this mass splitting for left-handed
leptons of the
$SU(2)\otimes U(1)$ model for $T>250$GeV and in a CP-asymmetric
medium.

\section{Computation of the self-energy}

Let us consider a non-abelian gauge field theory in which the
fermions, gauge bosons and scalars
do not have masses \cite{Weldon}.
We are interested in calculating the fermion
self-energy at the one-loop level at finite temperature in the
presence of a $CP$-asymmetric
fermionic background. We assume the chemical potentials $\mu_{f_i}$
of the different fermion species $f_i$ in the theory to be the same
$$\mu_{f_1}=\mu_{f_2}=\cdots=\mu_{f_n}=\mu\ .
\eqno(2.1)$$
We use the real-time formulation of finite-temperature
field theory and perform the calculation in the Feynman gauge. The
three relevant Feynman diagrams are shown in Fig. 1.

At tree level the free-particle (massless) propagators for fermion, gauge
boson and scalar in the Feynman gauge are given by
$$S(p)=\gamma_\mu p^{\mu}\left[{1\over p^2+i\epsilon}+i\Gamma_f (p)\right]\,,
\eqno(2.2)$$
$$D_{\mu \nu}(p)=-\eta_{\mu \nu}\left[{1\over p^2+i\epsilon}-i\Gamma_b (p)
\,,\right]
\eqno(2.3)$$
$$D(p)={1\over p^2+i\epsilon}-i\Gamma_b (p)\,,
\eqno(2.4)$$
where $\Gamma_f (p)$ and $\Gamma_b (p)$ are defined as
$$\Gamma_b (p)\equiv 2\pi \delta (p^2) n_b(p)
\eqno(2.5)$$
$$\Gamma_f (p)\equiv 2\pi \delta (p^2) n_f(p)
\eqno(2.6)$$
with
$$n_b(p)=\left[ e^{|p \cdot u |/T} -1 \right]^{-1}
\eqno(2.7)$$
$$n_f(p)=\theta(p\cdot u) n_f^{\scriptscriptstyle{-}}(p)
+\theta(-p\cdot u) n_f^{\scriptscriptstyle{+}}(p)\,.
\eqno(2.8)$$
In Eq. (2.8) the two functions $n_f^{\scriptscriptstyle{\mp}}(p)$
(not to be confused with the fermion number
density $N_f$ that will be introduced later) refer, respectively, to
fermions and anti-fermions. These differ from
$n_f(p)$ in the Weldon's paper in that they contain the chemical
potential $\mu_f$ of the fermion $f$
$$n_f^{\scriptscriptstyle{\mp}}(p)
= \left[ e^{(|p \cdot u | \mp \mu_f)/T} +1 \right]^{-1}\, .
\eqno(2.9)$$
The chemical potential $\mu_f$ is related to the number densities of fermions
$N_f$ and anti-fermions $N_{\bar f}$, and
is given , in the case of $\mu_f \ll T$, by
$$N_f-N_{\bar{f}}\equiv \int {d^3p\over(2\pi)^3}\left[
{n_f^{\scriptscriptstyle{-}}(p)-
n_f^{\scriptscriptstyle{+}}(p)}\right]\simeq {\mu_f T^2\over 6}\ .
\eqno(2.10)$$

We first consider the
tadpole diagrams: if all the particles in the theory are massless at
tree-level and if $\mu_{f_1}=\mu_{f_2}=\cdots=\mu_{f_n}$ then the
sum of all the relevant tadpole diagrams will be proportional to
$$\sum_m L^A_{mm}=0\ ,
\eqno(2.11)$$
where $L^A_{mm}$ are the representation matrices of the group
generators, $A$ runs over the generators and the two lower indices run
over the states of the fermion representation \cite{Weldon}. Therefore we
can see that the contributions from all the tadpole diagrams add up to
zero.

Since the calculations of the two diagrams in Fig. 1 are very similar,
we only will show some details of the calculation of the gauge boson
contribution to the one-loop fermion self-energy. Taking $\mu_{f_1}
=\mu_{f_2}=\cdots=\mu_{f_n}=\mu$ the sum of the gauge bosons
contribution is given by
$$\Sigma(K)=ig^2C(R) \int {d^4p\over(2\pi)^4} D_{\mu \nu}(p) \gamma^{\mu}
S(p+K)\gamma^{\nu}\ ,
\eqno(2.12)$$
where $g$ is the coupling constant and $C(R)$ is the quadratic Casimir
invariant of the gauge group representation.
We are only interested in the real part \cite{Weldon} of the finite temperature
corrections $\Sigma'$ to the fermion self-energy
$$\Sigma'= \Sigma-\Sigma\,(T=0)\,\,\,.
\eqno(2.13)$$
Substituting the expressions for the propagators Eqs. (2.2) and (2.3)
into Eq. (2.13), we obtain
$$Re\Sigma'=2g^2C(R)\int {d^4p\over(2\pi)^4} \left[
(\gamma_\mu p^{\mu} + \gamma_\mu K^{\mu})
{\Gamma_b (p) \over (p
+K)^2} - (\gamma_\mu p^{\mu} + \gamma_\mu K^{\mu})
{\Gamma_f (p+K)\over p^2} \right]\ .
\eqno(2.14)$$
Since Eq. (2.14) is manifestly covariant, it must be a linear combination
of $\gamma_\mu K^{\mu}$ and $\gamma_\mu u^{\mu}$. Following the procedure of
the Weldon's paper \cite{Weldon}, we obtain,
after performing the integration
over $p_0$ and over the angles,
$${1\over 4}Tr(\gamma_\mu K^{\mu}Re\Sigma')=g^2C(R)\int {dp\over8 \pi^2}
 \biggl\{ \biggl[ 4p- {K^2\over 2k
}L_1(p)\biggl]n_b(p) +$$
$$+\biggl[ 2p+ {K^2\over 2k }\left( \ln \left[{p+
\omega_{\scriptscriptstyle{+}} \over p+ \omega_{\scriptscriptstyle{-}}}
\right] - \ln \left[{ \omega_{\scriptscriptstyle{+}} \over
\omega_{\scriptscriptstyle{-}}} \right] \right)
\biggr] n_f^{\scriptscriptstyle{-}}(p) +$$
$$+\biggl[ 2p+ {K^2\over 2k }\left( - \ln \left[{p-
\omega_{\scriptscriptstyle{+}} \over p- \omega_{\scriptscriptstyle{-}}}
\right] + \ln \left[{ \omega_{\scriptscriptstyle{+}} \over
\omega_{\scriptscriptstyle{-}}} \right]
\right)\biggr] n_f^{\scriptscriptstyle{+}}(p) \biggr\}\ ,
\eqno(2.15)$$
$${1\over 4}Tr(\gamma_\mu u^{\mu}Re\Sigma')={g^2C(R)\over k }\int
{dp\over8 \pi^2} \biggl[ \left( 2p
\ln \left[{ \omega_{\scriptscriptstyle{+}} \over
\omega_{\scriptscriptstyle{-}}} \right] -pL_2(p)-
\omega L_1(p)\right)  {n_b(p) }$$
$$-p\left( \ln \left[{p+ \omega_{\scriptscriptstyle{+}} \over p+
\omega_{\scriptscriptstyle{-}}} \right]- \ln \left[{
\omega_{\scriptscriptstyle{+}} \over  \omega_{\scriptscriptstyle{-}}}
\right] \right) n_f^{\scriptscriptstyle{-}}(p)
-p\left( \ln \left[{p- \omega_{\scriptscriptstyle{+}} \over p-
\omega_{\scriptscriptstyle{-}}} \right]- \ln \left[{
\omega_{\scriptscriptstyle{+}} \over  \omega_{\scriptscriptstyle{-}}}
\right] \right) n_f^{\scriptscriptstyle{+}}(p)
\biggr]\ ,
\eqno(2.16)$$
where
$$\omega_{\scriptscriptstyle{\pm}}={1\over 2}(\omega \pm k)$$
$$L_1(p)=\ln \left[{p+ \omega_{\scriptscriptstyle{+}} \over p+
\omega_{\scriptscriptstyle{-}}} \right]-\ln \left[{p-
\omega_{\scriptscriptstyle{+}} \over p- \omega_{\scriptscriptstyle{-}}}
\right]$$
$$L_2(p)=\ln \left[{p+ \omega_{\scriptscriptstyle{+}} \over p+
\omega_{\scriptscriptstyle{-}}} \right]+\ln \left[{p-
\omega_{\scriptscriptstyle{+}} \over p- \omega_{\scriptscriptstyle{-}}} \right]
\ .
$$
Here all logarithms are to be understood in the principal-value
sense.

\section{Calculation of the effective mass splitting}

The two quantities in Eqs. (2.15) and (2.16) can be written in terms of
the Lorentz-invariants $a$ and $b$
$${1\over 4}Tr(\gamma_\mu K^{\mu}Re\Sigma')=-aK^2-bK\cdot u=
-a(\omega^2-k^2)-b\omega
\eqno(3.1)$$
$${1\over 4}Tr(\gamma_\mu u^{\mu}Re\Sigma')=-aK\cdot u-b=-a\omega-b\ ,
\eqno(3.2)$$
from which we obtain
$$ak^2={1\over 4}Tr(\gamma_\mu K^{\mu}Re\Sigma')-\omega{1\over 4}
Tr(\gamma_\mu u^{\mu}Re\Sigma')
\eqno(3.3)$$
$$bk^2=-\omega {1\over 4}Tr(\gamma_\mu K^{\mu}Re\Sigma')+(\omega^2-k^2)
{1\over 4}Tr(\gamma_\mu u^{\mu}Re\Sigma')\,\, .
\eqno(3.4)$$

The poles in the propagator occur, as can be seen from Eq. (1.3),
when $\omega$ and $k$ are such
as to produce a zero in the following Lorentz-invariant function
$$D=(1+a)^2K^2+b^2+2(1+a)bK\cdot u\, .
\eqno(3.5)$$
The positive-energy root occurs at
$$\omega(1+a)+b=k(1+a)\ .
\eqno(3.6)$$
We want to find the solution of this equation for $k=0$. We will call
$M$ the value of $\omega$ that satisfies Eq. (3.6) for $k=0$:
$$\omega(k=0)=M
\eqno(3.7)$$
and $M$ can be interpreted as an effective
fermion mass.

For small $k$, the two expressions ${1\over 4}Tr(\gamma_\mu K^{\mu}Re\Sigma')$
and ${1\over 4}Tr(\gamma_\mu u^{\mu}Re\Sigma')$ are of the form
$${1\over 4}Tr(\gamma_\mu K^{\mu}Re\Sigma')=h_0+h_1k^2+h_2k^4+\cdots
\eqno(3.9)$$
$${1\over 4}Tr(\gamma_\mu u^{\mu}Re\Sigma')=g_0+g_1k^2+g_2k^4+\cdots\ ,
\eqno(3.10)$$
where the $h_i$ and $g_i$ are functions of $\omega$ and $T$ alone.

We now want to use the expressions given by Eqs. (3.9) and
(3.10) to find the expression of $a$ and  $b$ for small $k$.
Inserting Eq. (3.9) and (3.10) into (3.3) and (3.4), we get
$$ak^2=h_0+h_1k^2-\omega g_0- \omega g_1k^2+\cdots
\eqno(3.11)$$
$$bk^2=-\omega h_0-\omega h_1k^2+(\omega^2-
k^2)(g_0+g_1k^2)+\cdots\ .
\eqno(3.12)$$
Since we have
$$h_0-\omega g_0=0\ ,
\eqno(3.13)$$
we obtain
$$a\simeq h_1- \omega g_1
\eqno(3.14)$$
$$b\simeq-\omega h_1+\omega^2g_1-g_0\ .
\eqno(3.15)$$
Using now Eqs. (3.14) and (3.15), we can rewrite Eq. (3.6)
for small $k$ as
$$\omega(1+h_1- \omega g_1)-\omega h_1+\omega^2g_1-
g_0=k(1+h_1- \omega g_1)
\eqno(3.16)$$
which, after simplification, becomes
$$\omega-k-g_0=k(h_1-\omega g_1)
\eqno(3.17)$$
and for $k=0$ gives
$$M=g_0
\eqno(3.18)$$
and therefore
$$M^2=h_0\ .
\eqno(3.19)$$

The quantity $h_0$ is given, from Eq. (3.9), by
$$h_0=\lim_{k\rightarrow 0}{1\over 4}Tr(\gamma_\mu K^{\mu}Re\Sigma')\ .
\eqno(3.20)$$
%
%
%%%%%%%%%%%%%%%%%%%%%%%%%%%%%%%%%%%%%%%
After a lengthy but straightforward algebra, Eq.(3.20) yields
$$
	h_0={g^2C(R)\over 8\pi^2}\biggl\{2\left({T^2 \pi^2 \over
	3}+{T^2 \pi^2 \over 6}+{\mu^2 \over 2}\right)-M \mu-
	M^2\int dp\left({1 \over p+M}-{1\over M -
	p}\right)n_b(p)
$$
$$
	+M^2\int dp\left[{1\over
	2p+M}\left(n_f^{\scriptscriptstyle{-}}(p)
        -{1\over e^{p/T}+1}\right)-{1\over
	M -2p}\left(n_f^{\scriptscriptstyle{+}}(p)
        -{1\over e^{p/T}+1}\right)
        \right]\biggr\}\, .
\eqno(3.21)
$$
Therefore, in order to find the effective
fermion  mass, we have to solve
the following equation for $M$
$$M^2=M_0^2\biggl\{
1+\left({\mu\over \pi T}\right)^2 -{M \mu \over \pi^2 T^2}-{M^2
\over\pi^2 T^2}\int_0^\infty dp\left({1
\over p+{M \over T}}-{1\over {M \over T} -
p}\right){1\over e^{p}-1}$$
$$
	+{M^2 \over\pi^2 T^2}\int dp\left[{1\over
	2p+M}\left({1\over e^{(p-\mu)/T}+1}
-{1\over e^{p/T}+1}\right)-{1\over
	M -2p}\left({1\over e^{(p+\mu)/T}+1}
-{1\over e^{p/T}+1}\right)
\right]\biggr\}\ ,
\eqno(3.22)$$
%%%%%%%%%%%%%%%%%%%%%%%%%%%%%%%%%%%%%%%%%%%%%%%%%
%
%
%
%
where
$$M_0^2=g^2{C(R)\over 8}T^2
\eqno(3.23)$$
is the value of the fermion effective mass in the absence of a
chemical potential \cite{Weldon}.
When we also add the scalar contributions to the fermion self-energy,
$M_0^2$ is modified to
$$M_0^2=g^2{C(R)\over 8}T^2+{C'|f|^2\over 16}T^2\ ,
\eqno(3.24)$$
where $|f|$ is the Yukawa coupling constant of the fermions and $C'$
is a numerical constant that depends on the fermion representations
\cite{Weldon}.
Equation (3.22) reduces to the result in Ref.[1] in the limit $\mu\to 0$,
as it should.

%
%
%
%
%%%%%%%%%%%%%%%%%%%%%%%%%%%%%%%%%%%%%%%
We now solve Eq.(3.22) for $M$ in the limit $\mu\ll T$. In this limit the
second line of Eq. (3.22) becomes
$$
	{M^2 \over\pi^2 T^2}\int dp\left[{1\over
	2p+M}\left({1\over e^{(p-\mu)/T}+1}
-{1\over e^{p/T}+1}\right)-{1\over
	M -2p}\left({1\over e^{(p+\mu)/T}+1}
-{1\over e^{p/T}+1}\right)\right]\simeq$$
$$
	{M^2 \over\pi^2 T^2}{\mu\over T}\int dp\left({1\over
	2p+M}+{1\over
	M -2p}\right){1\over e^{p/T}+1}\left(1-
	{1\over e^{p/T}+1}\right)\,.
\eqno(3.25)$$
Defining
$$
M\equiv M_0+x,
\eqno(3.26)
$$
with
$$
x \ll M_0,
\eqno(3.27)
$$
and neglecting higher order terms (such as the one in Eq. (3.25) ) and the
contribution from the  Yukawa couplings (i.e. $|f|^2\ll g^2$),
we obtain
$$x\simeq{M_0\over 2}\left[-g \sqrt{C(R)
\over 8\pi^4}{\mu\over T}-g^2{C(R)
\over 8\pi^2}\int_0^\infty dp\left({1
\over p+g\sqrt{C(R)
\over 8}}-{1\over g\sqrt{C(R)
\over 8}-
p}\right){1\over e^{p}-1}\right]\ .
\eqno(3.28)$$
In general we have
$${\mu \over T}\sim g^n\ ,$$
where the integer $n$ will depend on the specific gauge theory
and temperature range of interest,
therefore, using also Eq. (3.23), we can see that the value of the
fermion effective mass  is given, up to order $g^{n+1}$, by
$$M=M_0(1+c_1g+c_2g^2+\cdots +c_ng^n+ c_{n+1}g^{n+1}-
{g \over
2}{\sqrt{C(R) \over 8}}{\mu \over T}+\cdots)\ .
\eqno(3.29)$$
In order to find the values of the
$c_i$ ( which
are independent of $\mu$ and $T$) we would need to solve
numerically the integral of Eq. (3.28) and to include higher order
corrections to
the fermion self-energy (two-loops etc.).
The effective mass $\overline{M}$ of the antifermions can be
obtained by replacing $\mu$ by $-\mu$ in $M$.
Therefore, we find
$$
\Delta M \equiv \overline{M}-M=g^2 {C(R)\over 8\pi^2}\mu
	=\frac{3}{4 \pi^2}\frac{g^2C(R)}{T^2}\left(N_f-N_{\bar{f}}\right)\ .
\eqno(3.30)$$

\section{Discussion}

We have calculated the effective mass splitting between fermions and
anti-fermions in the CP-asymmetric background at high temperature to be
$\Delta M=\overline{M}-M=g^2 {C(R)\over 8\pi^2}\mu$.
Now, let us consider the case of $SU(2) \otimes U(1)$ at $T>250$GeV (when the
gauge symmetry is still unbroken) and with a $CP$-asymmetric
fermionic background.
Concentrating on the lepton sector of the first generation,
we can take the chemical
potential $\mu_{e_L}$ of the left-handed electrons,
$\mu_{e_R}$ of the right-handed electrons and the chemical potential
$\mu_\nu$ of the neutrinos to be the same
$$\mu_{e_L}=\mu_{e_R}=\mu_\nu=\mu\ .
\eqno(4.1)$$
(As pointed out in Ref.\cite{Barrchivukulafarhi},
if one considers only one generation of weak doublet,
due to the effect of electroweak fermion number violation
$N_f-N_{\bar{f}}$ would become zero.
However, with more than one doublet,
the difference between any two fermion numbers is conserved and the equilibrium
number density $N_f-N_{\bar{f}}$ at a given temperature for a given
generation will not be zero.)

In this case the mass splitting $\Delta
M$ between a left-handed anti-lepton and its anti-particle becomes
$$\Delta M={g'^2+3g^2\over 32\pi^2}\mu\ ,
\eqno(4.2)$$
where $g'$ is the coupling constant of the $U(1)$ field and $g$ is the
coupling constant of the $SU(2)$ gauge fields.
In deriving Eq.(4.2) we have used $C(R)=\frac{3}{4}$ and 1 for $SU(2)$
and $U(1)$ gauge groups, respectively, and the fact that the left-handed
fermions couple to the $U(1)$ gauge field with the strength $g^\prime /2$.

Using Eq. (2.10) and
$$N_{e_L}-N_{\bar {e_L}}\simeq N_\nu-N_{\bar \nu}=N_f-N_{\bar f}$$
we can write
$$\Delta M={3\over
16\pi^2}(g'^2+3g^2){N_f-N_{\bar{f}}\over T^2}\ .
\eqno(4.3)$$

Taking, based on $B-L$ conservation,
$$
\frac{N_f-N_{\bar{f}}}{N_\gamma}\simeq \frac{N_B-N_{\overline{B}}}{N_\gamma}
\sim 10^{-9},
\eqno(4.4)
$$
we see that
$$
N_f-N_{\bar{f}} \ \sim\ 10^{-9}\ T^3\ .
\eqno(4.5)
$$
Therefore, it appears that
the value of $\Delta M$ we obtained in Eq. (4.3) is too small to play a
significant role in the evolution of the $N_f-N_{\bar{f}}$
and consequently $N_B-N_{\overline{B}}$ in the
early Universe.
Possible effects of Eq. (4.3) on neutrino oscillations in the early Universe
will be given elsewhere.

\acknowledgements

One of us (A.E.) wishes to thank the High Energy Theory Group of
the Johns Hopkins University for the hospitality extended to him
during his several visits.
This work was supported in part by the National Science Foundation.

\clearpage
\noindent
{\large Figure Caption}

\noindent FIG. 1.
The 3 diagrams relevant to the one-loop self-energy calculation;
(a) the tadpole,\\
(b) the gauge boson contribution and (c) the scalar contribution.

%\begin{picture}(400,350)(-30,0)
% \put(-30,0){\psfig{figure=tadpole.ps,height=15truecm,width=15truecm}}
%\end{picture}

\end{document}